\documentstyle[a4,12pt,amssymb]{article}
\input epsf
\begin{document}
\def\thefootnote{\fnsymbol{footnote}}
\baselineskip=20pt
{\center
{\bf \huge Algebraic approximations to bifurcation curves of limit
cycles for the 
Li\'enard equation\\}
\vspace{1 cm}
{\large Hector Giacomini
\footnote{email: giacomini@univ-tours.fr}
 and S\'ebastien Neukirch 
\footnote{email: seb@celfi.phys.univ-tours.fr}\\}
Laboratoire de Math\'ematiques
 et Physique Th\'eorique\\ C.N.R.S. UPRES A6083 \\
Facult\'e des Sciences et Techniques, Universit\'e de Tours\\
F-37200 Tours FRANCE\\
}
\vspace{2 cm}
{\center \section*{Abstract}}
{ \small
In this paper, we study the bifurcation of limit 
cycles in Li\'enard systems of the form 
$\frac{d x}{d t}=y-F(x) \quad , \quad \frac{d y}{d t}=-x$
, where $F(x)$ is an odd polynomial that contains, 
in general, several free parameters.
By using a method introduced in a previous paper,
 we obtain a sequence of algebraic approximations
 to the bifurcation sets, in the parameter space.
Each algebraic approximation represents an exact lower
bound to the bifurcation set.
This sequence seems to converge to the exact bifurcation
 set of the system. The method is non perturbative. It
 is not necessary to have a small or
a large parameter in order to obtain these results.
}

\vspace{5 cm}

{\bf PACS numbers : 05.45.+b , 02.30.Hq , 02.60.Lj , 03.20.+i

Key words : Bifurcation, Li\'enard equation, limit cycles.}
\clearpage
The Li\'enard equation \cite{lienard}~:
\begin{equation}
\frac{d^2 x}{d t^2}+f(x) \frac{d x}{d t} +x=0 \label{equa lienard}
\end{equation}
appears very often within several branches of science,
 such as physics, chemistry, electronics, biology, etc (see
\cite {strogatz,h et s} and references therein).

This equation can be written as a two-dimensional dynamical
 system which reads as follows~:
\begin{equation}
\frac{d x}{d t}=y-F(x) \quad , \quad \frac{d y}{d t}=-x 
\label{equa lienard ode}
\end{equation}
where $F(x)=\int_0^x f(\tau) d\tau$.\\
The most difficult problem connected with the study of
equation (\ref{equa lienard ode}) is the question of the number
and location of limit cycles.

In order to make progress with this problem, it is of fundamental
importance to control the bifurcations of limit cycles that can take place
when one or several parameters of the system are varied.
The word bifurcation is used to describe any sudden change that
occurs while parameters are being smoothly varied in any dynamical
system.
Connections with the theory of bifurcations penetrate all natural 
phenomena. The differential equations describing real physical systems
always contain parameters whose exact values are, as a rule, unknown.
If an equation modeling a physical system is structurally instable,
that is if the behavior of its solutions may change qualitatively
 through arbitrary small changes in its right-hand side, then it
 is necessary to understand which bifurcations of its phase portrait
may occur through changes of the parameters.\\

In this respect, the most difficult bifurcation is the so-called saddle-node
bifurcation of limit cycles~:
let us suppose that system (\ref{equa lienard ode}) depends on
a parameter $\lambda$~: $F=F(x,\lambda)$.
Let $\Gamma_0$ be a non-hyperbolic limit cycle of
 (\ref{equa lienard ode}) (see \cite{perko2} for a definition)
, corresponding to the value $\lambda_0$ of the parameter $\lambda$.
System (\ref{equa lienard ode}) undergoes a saddle-node bifurcation at
$\lambda=\lambda_0$ if for $\lambda=\lambda_0+\epsilon$ and $\epsilon$
positive and sufficiently small, the limit cycle $\Gamma_0$ bifurcates
 into two hyperbolic limit cycles, one stable and the other instable.
 Moreover,  for $\lambda=\lambda_0-\epsilon$, the limit cycle $\Gamma_0$
disappears and there is no limit cycle in a small neighborhood of $\Gamma_0$.
This bifurcation is particularly difficult to detect because for
  $\lambda=\lambda_0-\epsilon$  there is no trace of it.
Moreover, the value of $\lambda_0$ is not known in principle and it is not
possible to employ a perturbative method with respect to $\epsilon$ to study
this type of bifurcation.

In a previous paper \cite{h et s}, we have introduced a method for studying
the number and location of limit cycles of (\ref{equa lienard ode}),
for the case where $F(x)$ is an odd polynomial of arbitrary degree.
The method is as follows~: we consider a function $h_n(x,y)$ given by~:
\begin{equation}
h_n(x,y)=y^n+g_{n-1,n}(x) y^{n-1}+g_{n-2,n}(x) y^{n-2}
+...+g_{1,n}(x) y + g_{0,n}(x) \label{hn}
\end{equation}
where $g_{j,n}(x)$, with $j=0,1,...,n-1$, are functions of only $x$
and $n$ is an even integer. Then it is always possible to choose
 the functions
$g_{j,n}(x)$ in such a way that~:
$$\dot{h}_n(x,y)=(y-F(x)) \frac{\partial h_n}{\partial x}-x 
\frac{\partial h_n}{\partial y}$$
is a function only of the variable $x$ (see also \cite{cherkas}).
 Then we have~:
\begin{equation}
\dot{h}_n(x,y)=R_n(x) \label{rn}
\end{equation}
The functions $g_{j,n}(x)$ and $R_n(x)$ determined in this way are
polynomials.
As explained in \cite{h et s}, if for a given value of $n$, the polynomial
$R_n(x)$ has no real roots of odd multiplicity, then the system has no
limit cycle.\\

We want to show in this paper that the method presented in 
\cite{h et s} enables us to determine algebraic approximations
to the bifurcation sets of limit cycles for the Li\'enard
equation. These bifurcation sets can be determined analytically only when the
system has a small parameter or a large one (perturbative regime).
In the intermediate case (non-perturbative regime), no method is known
for determining, in an analytic way, the bifurcation set.
We shall show here that our method gives a sequence of algebraic lower
bounds to the bifurcation sets. Moreover, this sequence seems to
converge to the exact bifurcation set.
The method can be applied to any system (\ref{equa lienard ode}) where $F(x)$
is an odd polynomial.\\

 As an example, we will consider a Rychkov system~:
\begin{equation}
F(x)=a_2 x^5+ a_1 x^3+ a_0 x \label{rych gene}
\end{equation}
with $a_2 \neq 0$.\\ 
We can take one of the parameters equal to one without loss of generality.
Several authors have studied this system with $F(x)$ written as
$F(x)=\epsilon (x^5-\mu x^3+x)$.
Rychkov has shown in \cite{rychkov} that this system
can have at most two limit cycles and actually has
 exactly two limit cycles when $\epsilon>0$ and $\mu>2.5$.
Rychkov's results have been improved
by Alsholm \cite{alsholm}, who lowered the bound of $\mu$ to $2.3178$ and
 by Odani \cite{odani}, who 
obtained an even smaller value $\sqrt{5}$.
By a scaling of the variables $x$ and $y$, system
(\ref{equa lienard ode}), with $F(x)$ given by (\ref{rych gene}), can
 be written in a more
simple form, as follows~:
\begin{equation}
F(x)=x^5-\mu x^3+\delta x \label{rych new}
\end{equation}
Since there are two parameters, the bifurcation set is given
 by a curve in the 
parameter plane $(\mu,\delta)$. Our aim, here, is to obtain 
information about the bifurcation
diagram of the system in this plane.
\begin{itemize}
\item
For $\delta<0$, thanks to Li\'enard theorem (see \cite{strogatz}),
 we know that the system 
has exactly one limit cycle for arbitrary values of $\mu$.
\item
For $\delta>0$ and $\mu<0$, the Bendixon criterium 
(see \cite{strogatz}) enables us to conclude
that the system has no limit cycle (the divergence of the vector
field, given by $-F'(x)$, has a constant sign for all x).
\item
For $\delta>0$ and $\mu>0$, the system can have two or zero limit cycles
, according to Rychkov's results. In this
region of the parameter space there exists a bifurcation curve
 $B(\mu,\delta)=0$. In the region $B(\mu,\delta)>0$, the system has exactly
two limit cycles and for $B(\mu,\delta)<0$, the system has no limit cycle.
On the curve $B(\mu,\delta)=0$ the system undergoes a saddle-node 
bifurcation~: there is a unique non-hyperbolic (double) limit cycle.
\end{itemize}

Obviously, the function $B(\mu,\delta)$ is not known and no analytical
method for obtaining this function for
arbitrary $\mu$ and $\delta$ exists. We shall obtain
a sequence of algebraic approximations to the function $B(\mu,\delta)$.

For a given even value of $n$, let us consider the corresponding polynomial
$R_n(x)$. The polynomials $R_n(x)$ described above, can have, for system
(\ref{rych new}), one, two or zero positive simple roots,
 depending on the values of $\mu$ and $\delta$.
 At least it is the behavior observed for the values of $n$ we considered.
\begin{itemize}
\item
For $\delta<0$ and $\forall \mu$ , the polynomials $R_n(x)$ have
one simple positive root.
\item
 For $\mu>0$ and $\delta > 0$ , the first quadrant is divided in two
 regions by a curve $B_n(\mu,\delta)=0$. In the region
 $B_n(\mu,\delta)>0$, the polynomial $R_n(x)$ has two positive simple
roots while in the region $B_n(\mu,\delta)<0$ it has no positive 
root. On the curve $B_n(\mu,\delta)=0$, $R_n(x)$ has a double positive root.
\item
For $\mu<0$ and $\delta > 0$, the polynomials $R_n(x)$ have no real root
other than the even-multiplicity root in $x=0$.
\end{itemize}
It is clear (see \cite{h et s}) that for $\mu>0$ and $\delta>0$ lying
 in the region
 $B_n(\mu,\delta)<0$, the system (\ref{equa lienard ode}) with $F(x)$ given
by (\ref{rych new}) has no limit cycle.
Hence, it is evident that the curve $B_n(\mu,\delta)=0$ represents an exact
 lower bound to the bifurcation curve $B(\mu,\delta)=0$~: the curves  
$B_n(\mu,\delta)=0$ are contained in the region $B(\mu,\delta)<0$
for all even values of $n$.

\begin{figure}[h]
$$
\epsfxsize=8cm
\epsfbox{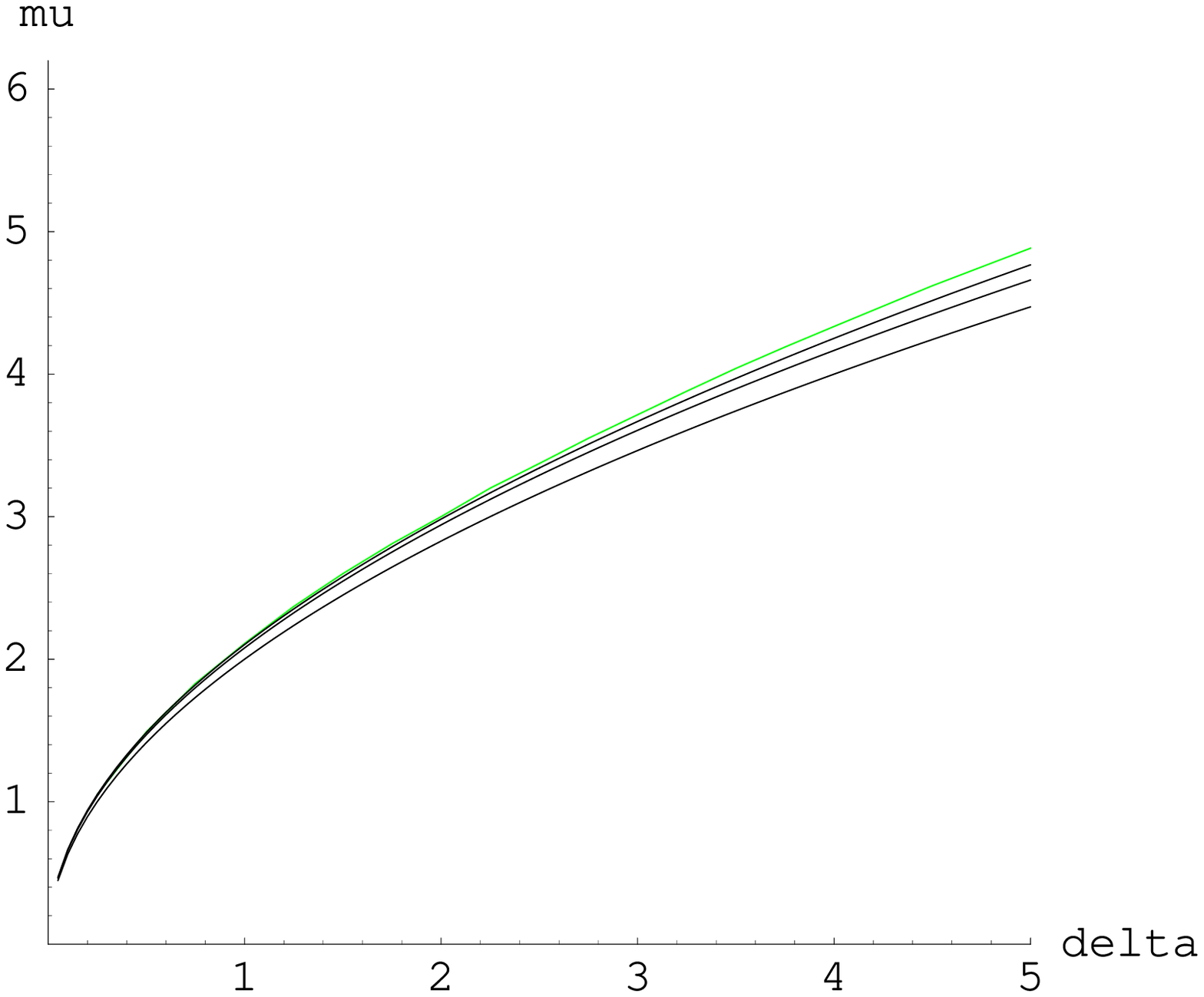}
$$
\caption{The algebraic curves $B_n(\mu,\delta)$
with $n=2,6,14$ (continuous lines) and the bifurcation 
curve $B(\mu,\delta)$ (dashed line) which is calculated
from numerical integrations of system (\ref{rych new}).}
\label{fig bif mu delta}
\end{figure}

\begin{figure}[htbp]
$$
\epsfxsize=8cm
\epsfbox{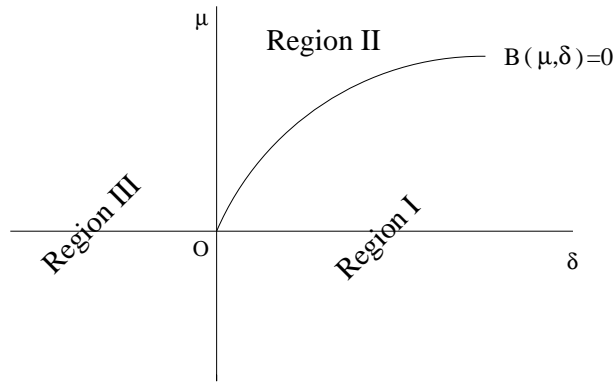}
$$
\caption{The complete bifurcation diagram of system 
(\ref{equa lienard ode}) with $F(x)$ given by (\ref{rych new}).
On the line $\delta=0$ the system undergoes a Hopf bifurcation.
On the curve $B(\mu,\delta)=0$ we have a saddle-node bifurcation
of limit cycles. The system has no limit cycle in region I, 
two limit cycles in region II and one limit cycle in region III.}
\label{fig zones}
\end{figure}

The functions $B_n(\mu,\delta)$ are algebraic and can be determined from the 
conditions~:
\begin{equation}
R_n(x)=0 \; \; \; \; \mbox{  and  } \; \; \; \; \frac{dR_n}{dx}(x)=0
\end{equation}
These two algebraic equations determine the double root of the polynomial
 $R_n(x)$ and give a relation between $\mu$ and $\delta$ which we write
 $B_n(\mu,\delta)=0$.
For $n=2$, we find $B_2(\mu,\delta)=\delta (\mu^2-4 \delta)$.
For $n=4$, $B_4(\mu,\delta)$ is a $12^{th}$ degree polynomial. The degree of
$B_n(\mu,\delta)$ increases rapidly with $n$.
We have calculated the functions $B_n(\mu,\delta)$ for even values of $n$
between 2 and 14.
The behavior of the curves $B_n(\mu,\delta)=0$, as well
 as the numerical bifurcation curve (calculated from a numerical integration
of the system), are shown in fig.(\ref{fig bif mu delta}). 
We see that each curve $B_n(\mu,\delta)=0$ is contained in the region
$B_{n-2}(\mu,\delta) > 0$. The complete bifurcation diagram is given 
in fig.(\ref{fig zones}). There are three regions~:
\begin{itemize}
\item[] Region I : There is no limit cycle. All the 
curves $B_n(\mu,\delta)=0$ lie in the part $\mu > 0$ of this region.
\item[] Region II : There are two limit cycles. All the polynomials
$R_n(x,\mu,\delta)$ have two positive simple roots.
\item[] Region III : Li\'enard theorem shows that there is one limit cycle.
All the polynomials $R_n(x,\mu,\delta)$ have one positive simple root.
\end{itemize}
We would like to emphasize that the shape of the bifurcation
curve $B(\mu,\delta)=0$ is already given by the curve 
$B_2(\mu,\delta)=\delta (\mu ^2-4 \delta)=0$, which is
constructed only with the function $F(x)$ ! The Hopf bifurcation happens when
$\delta=0$ and the saddle-node bifurcation occurs near $\mu^2=4 \delta$.  
\\

\begin{figure}[htbp]
$$
\epsfxsize=6cm
\epsfbox{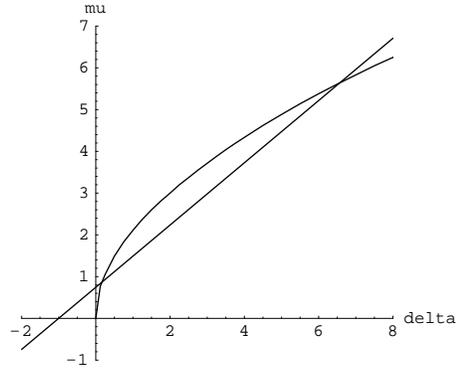}
$$
\caption{The line $\mu=\frac{\sqrt{5}}{3}(1+\delta)$ and the 
bifurcation curve $B(\mu,\delta)=0$ of system (\ref{rych new}).
The intersections between the two curves give the bifurcation
values of $\lambda$ for system (\ref{perko sys}).}
\label{fig ligne}
\end{figure}

We shall now make use of this bifurcation curve for the following system~:
\begin{eqnarray}
\dot{x} & = & y-(x^5-\frac{\sqrt{5}}{3} (1+\lambda) x^3 +\lambda x)
 \nonumber \\
\dot{y} & = & -x \label{perko sys}
\end{eqnarray}
This example has been studied by Lloyd \cite{lloyd2} and 
Perko \cite{perko2}. System (\ref{perko sys}) is a particular 
case of system (\ref{rych new}) with~: 
\[
\mu=\frac{\sqrt{5}}{3} (1+\lambda) \; \; \; \mbox{ and } \; \; \;
\delta=\lambda
\]
In order to know what are the bifurcations of this system
when the $\lambda$ parameter is varied from
$-\infty$ to $+\infty$, we must plot the line~:
\begin{equation}
\mu=\frac{\sqrt{5}}{3} (1+\delta) \label{eq ligne}
\end{equation}
in the bifurcation diagram of system (\ref{rych new}), with the
exact (but unknown) bifurcation curve $B(\mu,\delta)=0$ replaced by one of
the algebraic approximations $B_n(\mu,\delta)=0$.
As $\lambda$ is varied from $-\infty$ to $+\infty$,
the system moves along the line (\ref{eq ligne}) from left to right in 
fig.(\ref{fig ligne}).
It is easy to see that when $\lambda$ is negative (the portion of
the line is in region~III), there is one limit cycle. Then, when the line
 crosses the $\mu$ axis (that is when $\lambda$ changes sign) the system
 undergoes
a Hopf bifurcation~: a small limit cycle is created
around the origin of the phase plane. There are now two
 limit cycles, the system is in region~II. But when $\lambda$ is further
 increased, the line
 (\ref{eq ligne}) crosses the bifurcation curve $B(\mu,\delta)=0$~: the two
limit cycles collapse in a saddle-node bifurcation and there is no limit cycle
, the system is in region~I. If we continue to increase $\lambda$, we
 see the line (\ref{eq ligne}) crossing 
the curve $B(\mu,\delta)=0$ again~: two limit cycles appear in a saddle-node
bifurcation, the system enters region~II again. We can see that this is the
 last bifurcation we can create
because the line, when $\lambda$ is further increased, does not cross the
bifurcation set any more and stays in region~II. From the intersections
between the line (\ref{eq ligne}) and the curves $B_n(\mu,\delta)=0$, we
obtain algebraic approximations to the bifurcation values of the
parameter $\lambda$.\\

\begin{figure}[h]
$$
\epsfxsize=6cm
\epsfbox{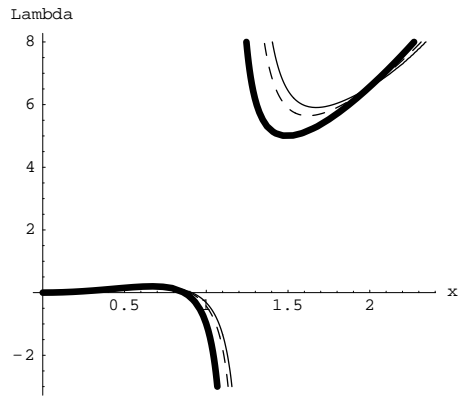}
$$
\caption{The algebraic approximations to the amplitude-bifurcation
diagram for system (\ref{perko sys}) for the cases $n=2$ (bold),
 $n=6$ (dash) and $n=10$ (continuous). We see the number and the
amplitudes of the limit cycles by considering the intersections
between one of these curves and a line $\lambda=\mbox{cte}$.
The results are improved with increasing values of $n$.}
\label{fig x-lambda}
\end{figure}

There is another way to see the different bifurcations of (\ref{perko sys}).
We have claimed in \cite{h et s} that if $n$ is large enough, the number
of positive roots of odd multiplicity of
$R_n(x)$ gives the number of limit cycles of the system.
In the case of system (\ref{perko sys}), $R_n=R_n(x,\lambda)$, so
 when $\lambda$ is varied, the number
of roots of $R_n$ changes. Hence, for a given value of $\lambda$, the
number of limit cycles of (\ref{perko sys})
can be obtained by counting the number of intersections between the curve
$R_n(x,\lambda)=0$ and the line $\lambda=\mbox{cte}$ in 
figure (\ref{fig x-lambda}).

\begin{figure}[htbp]
$$
\epsfxsize=6cm
\epsfbox{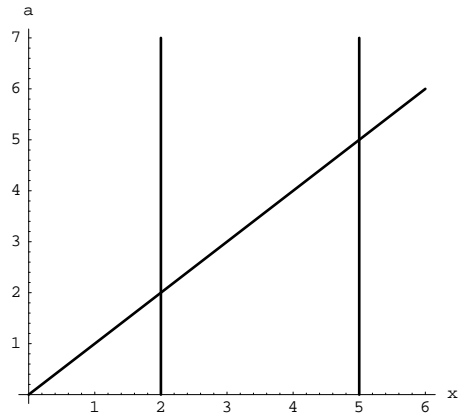}
$$
\caption{The curve $F(x,a)=0$ for system (\ref{eq 3 cycles}).
We see the number and the amplitude of the limit cycles by considering
the intersections between this curve and a line $a=\mbox{cte}$.
There seems to be transcritical bifurcations for $a=2$ and $a=5$ 
at this order.}
\label{fig 3 cycles 2}
\end{figure}

\begin{figure}[htbp]
$$
\epsfxsize=6cm
\epsfbox{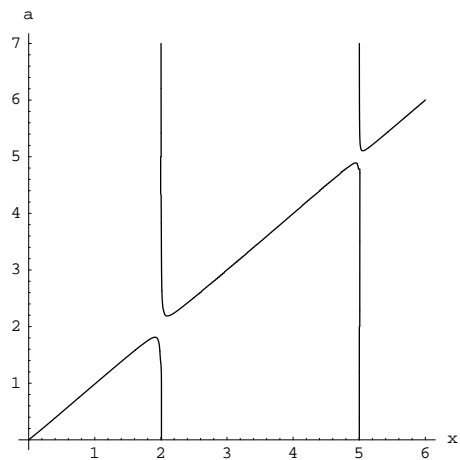}
$$
\caption{The curve $R_4(x,a)=0$ for system (\ref{eq 3 cycles}).
This curve is an approximation to the amplitude-bifurcation diagram.
We see the number of limit cycles by counting the number of intersections
between this curve and a line $a=\mbox{cte}$. The system only
presents saddle-node bifurcations. The qualitative behaviour of the
curves $R_n(x,a)=0$, with $n>4$, is the same. No further qualitative
changes occur for greater values of $n$.}
\label{fig 3 cycles 4}
\end{figure}

In \cite{h et s}, we claim that the value of the root of 
$R_n(x)$ gives an approximation to the maximum value of $x$ 
on the limit cycle (which we call the amplitude).
In fig.(\ref{fig x-lambda}) we have plotted $R_n(x,\lambda)=0$ 
for $n=2 \; , n=6$ and $n=10$. 
So we can see the amplitude of the limit cycles with respect
to $\lambda$. We see the Hopf bifurcation when $\lambda$ crosses
 the x-axis upward and we see the two saddle-node bifurcations
 when $R_n(x,\lambda)$ loses its two positive roots.
Once again, as $R_2(x,\lambda)=-2 x F(x,\lambda)$, an approximation to the
 bifurcation amplitude-diagram is given by the curve $F(x,\lambda)=0$, which
can be written as~:
$$
\lambda=x^2 \; \frac{3 x^2 - \sqrt{5}}{\sqrt{5} x^2 -3}
$$
\\
Let's consider another example~:
\begin{eqnarray}
\dot{x} & = & y- \left( x(x^2-a^2)(x^2-2^2)(x^2-5^2) \right) \nonumber \\
\dot{y} & = & -x \label{eq 3 cycles} 
\end{eqnarray}
We want to study the bifurcations of (\ref{eq 3 cycles}), when $a$ is
varied from $0 \to \infty$. As we already twice noticed, the qualitative
bifurcation amplitude-diagram seems to be given by $F(x)=0$. Here, the plot of
$F(x,a)=0$ seems to announce the presence of a transcritical bifurcation
(see \cite{strogatz} for a definition) near the values $a=2$ and $a=5$
 and a Hopf bifurcation for $a=0$ 
(see fig.(\ref{fig 3 cycles 2})).
If we plot $R_4(x,a)=0$, we still see the Hopf bifurcation, 
but the supposed-transcritical bifurcations are indeed  saddle-node
ones (see fig.(\ref{fig 3 cycles 4}))~: we see that the system can
have one or three limit cycles.
When $a$ is far from the values $a=2$ or $a=5$, there are three
 limit cycles,
but when $a$ is near the values $a=2$ or $a=5$, 
there is only one limit cycle. So, in
this example, the equation $F(x,a)=0$ does not give the right qualitative
amplitude-bifurcation diagram. We must plot the curve $R_4(x,a)=0$ in order
to obtain the good qualitative shape of it.\\

In summary, we have introduced a method that gives a sequence of
algebraic approximations to the bifurcation sets of limit cycles
for the Li\'enard equation (\ref{equa lienard ode}). These
algebraic approximations are exact lower bounds to the exact
bifurcation sets of the system and seem to converge to it in a monotonous
way. The fundamental aspect of this method is that it is
not perturbative in nature. It is not necessary to have a small
or a large parameter in order to apply it.\\

{\bf Acknowledgment~:} We thanks A. Gassul for bringing
reference \cite{cherkas} to our attention after we presented him
the results given in \cite{h et s}.

\end{document}